\documentclass[twocolumn,floats,floatfix,superscriptaddress,aps,pra]{revtex4}
\usepackage{amsmath}
\usepackage{graphicx}
\usepackage{multirow}
\usepackage{booktabs}
\usepackage{afterpage}

\usepackage{array}
\newcolumntype{L}[1]{>{\raggedright\let\newline\\\arraybackslash\hspace{0pt}}m{#1}}
\newcolumntype{C}[1]{>{\centering\let\newline\\\arraybackslash\hspace{0pt}}m{#1}}
\newcolumntype{R}[1]{>{\raggedleft\let\newline\\\arraybackslash\hspace{0pt}}m{#1}}

\def\be{ \begin{equation} }
\def\ee{ \end{equation} }
\def\H{\mathbf{H}}
\def\U{\mathbf{U}}
\def\O{\mathcal{O}}
\def\A{A}

\def\eps{\epsilon}
\def\error{\alpha}
\def\bw{R}
\def\params{\Delta_k, \Omega}

\def\matrix22#1#2#3#4{\left[ \begin{array}{cc} #1 & #2 \\ #3 & #4 \end{array}\right]}

\def\bt{\begin{tabular}}
\def\et{\end{tabular}}

\def\rms{\tilde{\Omega}}
\def\rms{\Lambda}
\def\area{A}

\def\bw{\eps_0}

\begin{document}

\author{Svetoslav S. Ivanov}
\affiliation{Department of Physics, St Kliment Ohridski University of Sofia, 5 James Bourchier Blvd, 1164 Sofia, Bulgaria}
\author{Boyan T. Torosov}
\affiliation{Institute of Solid State Physics, Bulgarian Academy of Sciences, 72 Tsarigradsko chauss\'{e}e, 1784 Sofia, Bulgaria}
\author{Nikolay V. Vitanov}
\affiliation{Department of Physics, St Kliment Ohridski University of Sofia, 5 James Bourchier Blvd, 1164 Sofia, Bulgaria}

\title{High-fidelity quantum control by polychromatic pulse trains}

\date{\today}

\begin{abstract}

We introduce a quantum control technique using polychromatic pulse sequences (PPS), consisting of pulses with different carrier frequencies, i.e. different detunings with respect to the qubit transition frequency. 
We derive numerous PPS, which generate broadband, narrowband, and passband excitation profiles for different target transition probabilities.
This makes it possible to create high-fidelity excitation profiles which are either (i) robust to deviations in the experimental parameters, 
which is attractive for quantum computing, or (ii) more sensitive to such variations, 
which is attractive for cross talk elimination and quantum sensing. 
The method is demonstrated experimentally using one of IBM's superconducting quantum processors, in a very good agreement between theory and experiment.
These results demonstrate both the excellent coherence properties of the IBM qubits and the accuracy, robustness and flexibility of the proposed quantum control technique.
They also show that the detuning is as efficient control parameter as the pulse phase that is commonly used in composite pulses.
Hence the method opens a variety of perspectives for quantum control in areas where phase manipulation is difficult or inaccurate.

\end{abstract}

\maketitle


\textbf{Introduction.} 
The composite pulses are a powerful quantum control technique which has been  invented in nuclear magnetic resonance (NMR) as a convenient tool for robust manipulation of spins by magnetic fields \cite{NMR, Levitt1986, Levitt2007}. A composite pulse consists of a sequence of single pulses each having a well defined phase. 
With a proper selection of the phases of the constituent pulses, the composite sequence can largely compensate for systematic errors in the driving fields (e.g. miscalibrated, shifted or spatially inhomogeneous intensity or frequency), which usually lead to poor performance of a single pulse alone. 
Moreover, composite pulses feature high accuracy in obtaining the targeted probability value, as well as great flexibility for they can produce an excitation profile of virtually any shape. 
After their successful application in NMR, composite pulses have been applied in other areas where robust coherent quantum control is needed.
Examples range from trapped ions \cite{Gulde2003, Schmidt-Kaler2003, Haffner2008, Timoney2008, Monz2009, Shappert2013, Mount2015, Zarantonello2019}, neutral atoms  \cite{Rakreungdet2009,Demeter2016}, quantum dots \cite{Wang2012,Kestner2013,Wang2014,Zhang2017,Hickman2013,Eng2015}, and NV centers in diamond \cite{Rong2015} to doped solids \cite{Schraft2013,Genov2017,Bruns2018,Genov2014}, superconducting phase qubits \cite{SteffenMartinisChuang},
 optical clocks \cite{Zanon-Willette2018}, atom optics \cite{Butts2013,Dunning2014,Berg2015}, and magnetometry \cite{Aiello2013}.

In this work, we construct composite pulse sequences in which, instead of the phase, we use the detuning of each constituent pulse as the control parameter, thereby forming a polychromatic pulse train (PPT).
We are motivated by the fact that in the vast library of traditional composite pulses the standard control parameters are the phases and, to some (but lesser) extent, the amplitudes of the constituent pulses.
In particular, the phases have proved to be a much more powerful control parameter than the pulse amplitudes.
Using the frequency of each pulse as a control parameter has largely been ignored, with just a few exceptions \cite{Torosov2019,Kyoseva2019}.
Here we explore how much one can achieve by using the detuning as the only control parameter by deriving, and experimentally demonstrating on IBM's ibmq\_armonk quantum processor, all major types of excitation profiles of the traditional composite pulses.
Remarkably, it turns out that PPT can deliver the same performance in terms of efficiency, robustness and flexibility as traditional composite pulses. 

\begin{figure}[tb]
	\includegraphics[width=0.8\columnwidth]{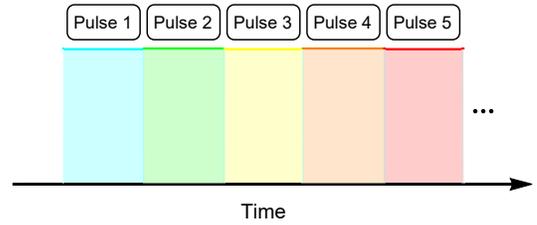}
	\caption{Schematic illustration of a polychromatic pulse train: a sequence of pulses with different frequencies. }
	\label{Fig:PulseTrain}
\end{figure}


\textbf{General framework.} The Hamiltonian of a coherently driven qubit can be written as  
$\H =  \frac{1}{2} \hbar \Omega \sigma_x - \frac{1}{2} \hbar \Delta \sigma_z$, where $\sigma_k$ are the Pauli matrices,
$\Delta=\omega_0-\omega$ is the detuning between the transition frequency $\omega_0$ and the frequency of the driving field $\omega$, and $\Omega$ is the Rabi frequency.
The evolution is described by a propagator, which can be parameterized with the complex Cayley-Klein parameters $a$ and $b$ as
\be\label{SU2}
\U = \matrix22{a}{b}{-b^{\ast}}{a^{\ast}},
\ee
where $|a|^2+|b|^2=1$.
For rectangular pulse of duration $T$ and constant detuning, which we assume below, the propagator is obtained by exponentiation, 
and we have
\begin{equation}
  a = \cos\frac{\area}{2} + i \frac{\Delta}{\rms}\sin \frac{\area}{2}, \qquad
  b = -i\frac{\Omega}{\rms}  \sin \frac{\area}{2},
\end{equation}
where $\rms = \sqrt{\Delta^2 + \Omega^2}$ and $\area = \rms T$.
The transition probability is $p = \left|b\right|^2 =(\Omega/\rms)^2\sin^2(\area/2)$.

The fastest way to drive complete population transfer, or to create a coherent superposition of states, is to use a resonant pulse ($\Delta=0$) with an appropriate temporal area, for which the transition probability is $p=\sin^2 \A/2$.
Obviously, it is sensitive to variations in the pulse area. 
For example, a small deviation $\eps$ from the value $\A=\pi$, i.e., $\A \rightarrow \pi(1 + \eps)$, causes an error in the transition probability of order $\O(\eps^2)$: $p = 1 - \pi^2\eps^2/4 + \ldots$.
In the more general case of fractional-$\pi$ pulses, 
the error in $p$ is $\O(\eps)$.
This sensitivity to errors can be greatly reduced, to any
desired order, by replacing the single pulse by a PPT.
A PPT of $N$ pulses of the same duration $T$, the same Rabi frequency $\Omega$ and the same phases, but with different detunings $\Delta_k$ (leftmost pulse applied first), as illustrated in Fig.~\ref{Fig:PulseTrain},
\be\label{sequence}
\{ \Delta_1 \Delta_2 \cdots \Delta_N \} ,
\ee
produces the propagator
\be\label{sequencePropagator}
\U^{(N)} = \U(\Delta_N) \cdots \U(\Delta_2) \U(\Delta_1),
\ee
which is a product of the single-pulse propagators \eqref{SU2}.
It depends on all detunings, which can be varied in order to engineer the excitation profile in a desired manner.
%


\textbf{Description of the experiment.} We performed our experiments using the quantum processor ibmq\_armonk v2.4.25 of IBM Quantum Experience \cite{ibm_quantum}. It consists of a single transmon qubit, which is controlled with microwave pulses, using the low-level quantum-computing module Qiskit Pulse \cite{QiskitPulse}, part of the open-source framework Qiskit \cite{Qiskit}. The system parameters, as calibrated at the time of the experiments, are: qubit frequency 4.972 GHz; anharmonicity $-0.34719$ GHz; T1 and T2 times 195.52 $\mu$s and 232.57 $\mu$s; readout assignment error 3.47\%. 
We apply sequences of rectangular pulses with the same drive amplitude (Rabi frequency), where each pulse has a duration of 100 ns. 
Each pulse has a different carrier frequency (and hence different detuning), as explained above. 
Each single experimental data point on the figures below represents the average of 1024 shots.

We note here that in the figures, presenting the experimental results, the excitation profiles do not reach the desired maximum value of unity, contrary to the theoretical predictions. 
This is largely due to the measurement error, which is on the order of 3.5\%. A much smaller but more important amount, is contributed from the dephasing of our qubit. 
The malign effect of the decoherence can be mitigated by shortening the pulse duration, which however results in higher leakage out of the computational space of the qubit
\footnote{This problem is a topic of a separate research, which will be published later.}.

\begin{figure}[tbph]
	\includegraphics[width=\columnwidth]{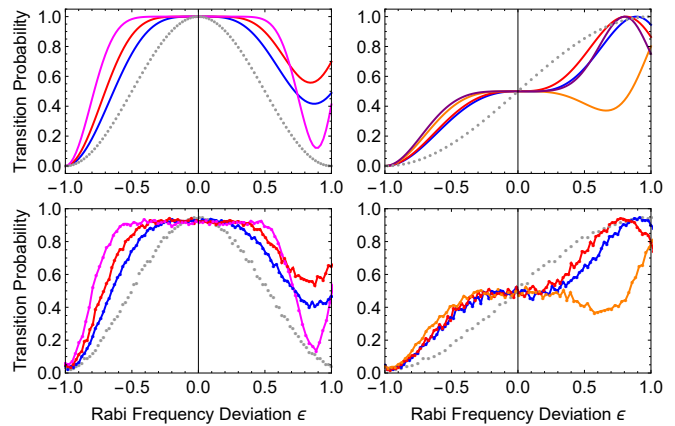}
	\caption{
	Transition probabilities generated by broadband antisymmetric PPTs of Eq.~\eqref{antisym sequence}.
	Top frames: numerical simulations; bottom frames: experimental results.
	\textit{Left frames}: 
	PPTs with length $N=3$, 5, and 11 (from inside out) for complete population transfer ($p=1$), compared to the profile of a single resonant pulse (dotted). 
    The values of the Rabi frequency and detunings ($\Omega;\Delta_1,\Delta_2,\ldots,\Delta_n$) are: (0.6397; 0.7200) for $N=3$; (0.5583; 0.8980, 0.1412) for $N=5$; and (0.4795; 1.1164, 0.2309, 0.4414, 0.0233, 0.1611) for $N=11$.
	\textit{Right frames}: 
	PPTs with length $N=3$, 5, and 7 (from inside out) for partial population transfer ($p=\frac12$), compared to a single resonant pulse (dotted). 
	The values of the Rabi frequency and detunings ($\Omega;\Delta_1,\Delta_2,\ldots,\Delta_n$) are (0.7014; 1.1789) for $N=3$, (0.4498; 0.8182, 0.4731) for $N=5$, and (0.4875; 1.0942, 0.2006, 0.5543) for $N=7$.}
	\label{Fig:BB-deriv}
\end{figure}

\textbf{Broadband (BB) pulses.} 
Although the propagator for every pulse in the sequence \eqref{sequencePropagator} is first-order sensitive to the pulse area error $\eps$,
the detunings $\Delta_k$ and the Rabi frequency $\Omega$ can be chosen such that in the PPT this sensitivity is relegated to much higher orders of $\eps$.
We begin with complete population inversion sequences.
We found that for transition probability of $p=1$, the best performance is obtained when we take an odd number of pulses $N=2n+1$ with the same Rabi frequency $\Omega$ and set the anti-symmetric condition $\Delta_{N-k+1} = -\Delta_k$ on the detunings.
Explicitly, this PPT reads
\be\label{antisym sequence}
\{\Delta_1,\Delta_2, \cdots ,\Delta_n, 0, -\Delta_n, \cdots, -\Delta_2, -\Delta_1 \}.
\ee

We make use of two approaches to derive broadband PPTs.
In the \textit{first} approach, we require that $p=1$ at the center of the excitation profile, i.e. for zero pulse area error, and we set as many derivatives  with respect to $\epsilon$ to zero as possible,
\begin{subequations}\label{deriv}
\begin{align}
P_N (\epsilon = 0) &= 1, \\
\left. \frac{\partial^k P_N }{\partial \epsilon^k} \right|_{\epsilon = 0}  &= 0  \quad (k=1,2,\ldots, n).
\end{align}
\end{subequations}
Hence the transition probability $P_N = 1 - |U_{11}^{(N)}|^2$ is accurate to order $O(\epsilon^{2n}) = O(\epsilon^{N-1})$.
The conditions \eqref{deriv} generate a set of $n+1$ algebraic equations for the $n+1$ parameters $\{\Omega; \Delta_1, \Delta_2, \ldots, \Delta_n\}$.
Contrary to the traditional composite pulses, where 
numerous analytic solutions for the phases have been derived, here an analytic solution is elusive even for the shortest PPTs due to algebraic complexity.
Numerical solutions, however, are straightforward.
Among them, we have selected the ones with the least total pulse area; they are listed in the Supplementary Material.

\begin{figure*}[tb]
\includegraphics[width=1.7\columnwidth]{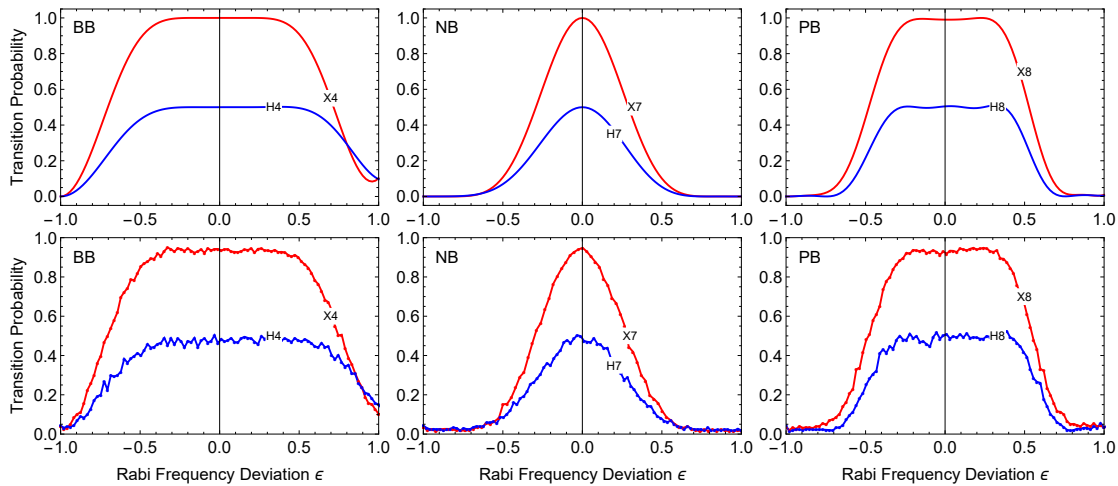} 
	\caption{
	\textit{Left}: Transition probabilities generated by broadband PPTs of length $N=4$. The curves show the transition probabilities locked at the levels $p = 1$ (X4) and $p = 0.5$ (H4). The probability error level is $\error=10^{-4}$. Top frame: numerical simulations; bottom frame: experimental results. The Rabi frequency and detunings ($\Omega;\Delta_1,\Delta_2,\Delta_3,\Delta_4$) are $(0.6750; -0.9267, 0.0227, -0.0227, 0.9267)$ for X4 and $(0.6197; 0.6465, -0.0024, -1.1049, 0.6265)$ for H4.
	\textit{Middle}: Transition probabilities generated by narrowband symmetric PPTs of length $N=7$, locked at the levels $p=1$ (X7) and $p=0.5$ (H7). The error level is $10^{-4}$. 
	Top frame: numerical simulations; bottom frame: experimental results. The Rabi frequency and detunings ($\Omega;\Delta_1,\Delta_2,\Delta_3,\Delta_4$) are (0.4036; 0.7207, $-0.1269$, 0.2682, 0.5699) for X7 and (0.3849; 0.0807, 0.3045, 0.7847, $-0.6154$) for H4.
	\textit{Right}: Transition probabilities generated by passband PPTs of length $N=8$ locked at the levels $p=1$ (X8, $\bw=0.2$) and $p=0.5$ (H8, $\bw=0.3$). The error level is $10^{-2}$. 
	Top frame: numerical simulations; bottom frame: experimental results. 
	The Rabi frequency and detunings ($\Omega;\Delta_1,\Delta_2,\ldots,\Delta_8$) are (0.6197; 0.3847, 0.5165, $-2.4852$, 0.549, 0.4073, 0.3837, 0.0138, -0.8335) for X8 and (0.825; 2.6171, 0.5036, 0.2977, 0.1954, 0.8605, $-0.6183$, 1.8844, 1.8191) for H8.
	}
	\label{Fig:num}
\end{figure*}

Figure \ref{Fig:BB-deriv} shows the excitation profiles generated by some of the BB PPTs for target transition probabilities $p=1$ and $p=\frac12$, respectively.
Note that a zero error $\epsilon=0$ corresponds to a different Rabi frequency $\Omega$ (which is a control parameter) for each sequence.
As the number of pulses in PPT grows, the excitation profiles broaden and become more robust to Rabi frequency errors.
We note that throughout this paper we set the pulse duration $T=1$ and all values of $\Omega$ and $\Delta_k$ are dimensionless; in other words, all values of $\Omega$ and $\Delta_k$ are in units $1/T$.

In the \textit{second} approach, the parameters $\Delta_k$ and $\Omega$ are obtained by a numerical minimization of the following cost (objective) function,
\begin{equation}
\label{costBB}
\text{cost}(\params) = \left[\max\left(\left|p(\params, \eps)-p\right|\right) - \error\right]^2,
\end{equation}
where $\eps$ is varied in the range from $-\bw$ to $\bw$, which defines the bandwidth of the broadband pulse.
For the optimization we use a BFGS quasi-Newton method.
We must note that optimizing the cost function \eqref{costBB} means that we would actually penalize transition profiles that do not utilize the allowed deviation range, specified by $\error$.
In other words we force the excitation profiles to be erroneous up to the quality level defined by $\error$, in exchange of enhanced bandwidth.

For the numeric calculations we vary $\eps$ with a step of 0.1, as we observe that a finer step only slows down the optimization. To validate a solution, we make sure that
$
\left|p(\params, \eps)-p\right| \leq \error
$
holds over the entire bandwidth $[-\bw,\bw]$.
Note that despite being optimized for a certain bandwidth $\bw$, the obtained PPTs often outperform that requirement: the above condition holds over a broader range for $\eps$.
As in the first approach, for $p=1$ the best performance is obtained with the antisymmetric PPT \eqref{antisym sequence}. 
For $p\neq1$, we use the general sequence \eqref{sequence}.

The full list of broadband sequences that we obtained can be found in the Supplementary Material or in the GitHub repository \cite{GitHub}.
Selected profiles are shown in Fig. \ref{Fig:num} (left) for $N=4$ pulses, $p=1$ (X4) and 0.5 (H4) at probability error level $\alpha=10^{-4}$. 
The advantage of these PPTs over the derivative-based ones in Fig.~\ref{Fig:BB-deriv} is the enhanced bandwidth.
Because the admissible error here is very small ($\alpha=10^{-4}$) the loss of accuracy is invisible on this scale. (The error of the derivative-based PPTs in their high-accuracy central range is generally extremely small, far less than $10^{-4}$.)


\textbf{Narrowband (NB) pulses.} We derive the NB pulses by minimizing the following cost function for $\Delta_k$ and $\Omega$,
\begin{multline}
\label{costNB}
\text{cost}(\params) =
\left[p(\params, 0) - p\right]^2 +
\left(\left.\frac{\partial p(\params, \eps)}{\partial \eps}\right|_{\eps=0}\right)^2 \\
+ \left[\max\left(p(\params, \eps)\right) - \error\right]^2,
\end{multline}
where $\eps$ sweeps the range $\left[-1,-\bw\right]\cup\left[\bw, 1\right]$.
Minimizing the derivative locks the extremum of the probability at $\eps=0$. For $p=1$ this derivative is zero and it can be omitted from Eq. \eqref{costNB}.
To validate a solution, we make sure that
$
\left|p(\params, 0) - p\right| \leq \error
$
is fulfilled and that
$
p(\params, \eps) \leq \error
$
holds over the entire bandwidth.

We found that the best performance is obtained when we set the symmetric condition $\Delta_{N-k+1} = \Delta_k$.
Then the PPT reads
\be
\{\Delta_1,\Delta_2, \cdots ,\Delta_n, 0, \Delta_n, \cdots, \Delta_2, \Delta_1 \}.
\ee
In Fig. \ref{Fig:num} (middle) we show selected profiles for $N=7$ and target probabilities $p=1$ (X7) and 0.5 (H7) at error level $\alpha=10^{-4}$. 
%
The full list of the obtained NB PPTs can be found in the Supplementary Material or in the GitHub repository \cite{GitHub}.


\textbf{Passband (PB) pulses.} We derive the PB sequences by numerical minimization of the following cost function of $\Delta_k$ and $\Omega$,
\begin{multline}
\label{costPB}
\text{cost}(\params) =
\left([\max (\left|p(\params, \eps_1)-p\right|)-\error\right]^2 \\
+\left[\max (p(\params, \eps_2))-\error\right]^2,
\end{multline}
where $\eps_1$ is varied in the range $\left[-\bw,\bw\right]$ and $\eps_2$ is varied in the range $\left[-1,-\bw\right]\cup\left[\bw, 1\right]$.
This is essentially a combination of the BB cost function of Eq. \eqref{costBB}, which enhances the flat nature in the middle (around $\eps=0$), and the NB cost function of Eq. \eqref{costNB}, which suppresses the wings of the profile. 
Note that now the first two terms in Eq. \eqref{costNB} are redundant.
To reduce the number of solutions that we obtain, both terms from Eq. \eqref{costPB} share the same bandwidths $\bw$ and errors $\error$.
To validate a solution, we make sure that the BB condition  holds over the range $\left[-\bw,\bw\right]$ and the NB condition holds over the range $\left[-1,-\bw\right]\cup\left[\bw, 1\right]$.
The full list of obtained passband sequences can be found in the Supplementary Material or in the GitHub repository \cite{GitHub}.

\begin{figure}[t!]
	\includegraphics[width=0.9\columnwidth]{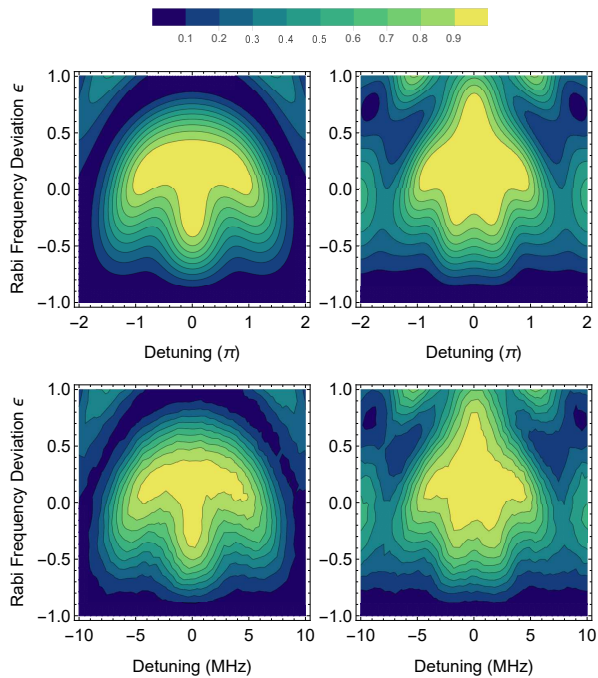}
	\caption{Contour plots of the transition probability generated by PPTs with double compensation of Rabi frequency (relative) and detuning (absolute) errors for $N=2$ (left frames) and $N=4$ (right frames). 
	Top frames: numerical simulations; bottom frames: experimental results. 
	The values of the parameters are $(\Omega; \Delta_1, \Delta_2) = (0.937; 0.735, -0.735)$ for $N=2$ and $(\Omega; \Delta_1, \Delta_2, \Delta_3, \Delta_4)=(0.9; 3.028, 0.609, -0.609, -3.028)$ for $N=4$.}
	\label{Fig:2D}
\end{figure}

In Fig. \ref{Fig:num} (right) we show selected profiles for $N=8$, target probabilities $p=1$ (X8) and 0.5 (H8) and probability error $\alpha=10^{-2}$.
%
The comparison of the BB, NB and PB profiles in Fig. \ref{Fig:num} demonstrates that the PB profiles maintain the NB feature of the NB profiles, however, with the added benefit of a flat BB top.
Therefore, the PB composite sequences feature both selectivity and robustness, at the expense of being longer than the BB and the NB sequences.

\textbf{Simultaneous compensation of Rabi frequency and detuning errors.} PPTs can be used to achieve robustness against various imperfection, and are not limited to Rabi frequency error compensation. We have used our approach to derive PPTs with double compensation versus both Rabi frequency and detuning.
The obtained sequences have an antisymmetric form and
the excitation profiles for $N=2$ and 4 pulses are plotted in Fig.~\ref{Fig:2D}. 
As seen in the figure, simultaneous compensation of Rabi frequency and detuning errors can be efficiently achieved. 
The broadening of the high-efficiency domain for 4 pulses is clearly visible.
Longer PPTs, not shown here for the sake of brevity, can further expand this domain. 
As with all previous PPTs, an excellent agreement between theory and experiment is observed.


\textbf{Discussion and conclusions.} 
We presented a method for robust and high-fidelity quantum control of qubits by sequences of nearly-resonant pulses with different, suitably chosen carrier frequencies (and hence detunings) used as control parameters. 
The method allowed us to derive sequences with broadband, narrowband, and passband excitaiton profiles, robust coherent superpositions, as well as sequences with double compensation of both Rabi frequency and detuning errors. 
We performed experimental tests of all sequences by using IBM's quantum processor ibmq\_armonk. 
All tests demonstrated an excellent agreement between theory and experiment.
Our results show that using the detuning as a control parameter is as efficient as using the pulse phases, as done in the vast field of quantum control by composite pulses.
This opens a variety of perspectives to benefit from the powerful composite ideas in areas where phase control is inaccurate, difficult, or even impossible.


\acknowledgments
This work is supported by the European Commission's Horizon-2020 Flagship on Quantum Technologies project 820314 (MicroQC). We acknowledge the use of IBM Quantum services for this work. The views expressed are those of the authors, and do not reflect the official policy or position of IBM or the IBM Quantum team.

\end{document}